\documentclass[aps,twocolumn,superscriptaddress,longbibliography]{revtex4-1}

\usepackage[colorlinks=true,citecolor=blue]{hyperref} 
\usepackage{graphicx}
\usepackage{bm}
\usepackage{amsmath}
\usepackage{amssymb}

\DeclareGraphicsExtensions{.png,.jpg,.eps}
\usepackage{xcolor}

\begin{document}

\author{Wonhee Ko}
\thanks{These authors contributed equally.}
\affiliation{Center for Nanophase Materials Sciences, Oak Ridge National Laboratory, Oak Ridge, Tennessee 37831, USA}
\affiliation{Department of Physics and Astronomy, University of Tennessee, Knoxville, Tennessee 37996, USA}

\author{Sang Yong Song}
\thanks{These authors contributed equally.}
\affiliation{Center for Nanophase Materials Sciences, Oak Ridge National Laboratory, Oak Ridge, Tennessee 37831, USA}

\author{Jiaqiang Yan}
\affiliation{Materials Science and Technology Division, Oak Ridge National Laboratory, Oak Ridge, Tennessee 37831, USA}

\author{Jose L. Lado}
\affiliation{Department of Applied Physics, Aalto University, 02150 Espoo, Finland}

\author{Petro Maksymovych}
\email{maksymovychp@ornl.gov}
\affiliation{Center for Nanophase Materials Sciences, Oak Ridge National Laboratory, Oak Ridge, Tennessee 37831, USA}

\title{
  Atomic scale imaging of sign-changing order parameter in superconducting FeSe with Tunneling Andreev Reflection
}

\begin{abstract}
The pairing symmetry of the superconducting order parameter in iron-based superconductors has been a subject of debate, with various models proposing $s$-wave,
$d$-wave, and mixed combinations as possible candidates.
Here we probe the pairing symmetry of FeSe utilizing the new methodology of Tunneling Andreev Reflection (TAR). TAR directly exploits the transparency-dependence of tunneling current to disentangle contributions of single-particle current and Andreev reflection. These measurements provided new direct evidence in favor of the sign-changing nature of the superconducting order in FeSe, in a distinctly complementary approach to nanoscale imaging
of quasiparticle interference. Crucially TAR can also probe higher-order contributions to Andreev reflection. Quantitative comparison of the experimental signatures of higher-order Andreev reflections with those in concomitant tight-binding simulations revealed new evidence in support of the nodal gap structure of superconductivity in FeSe. Finally, the effect of structural topological defects can be directly probed with TAR owing to its atomic spatial resolution. In particular, we find that superconductivity is completely suppressed along the twin boundary while its electronic structure is characterized by a V-shaped signature of a pseudogap state. Our findings provide new insight into the pairing symmetry of an unconventional superconductor, demonstrating the potential of differential tunneling Andreev reflection to reveal microscopic properties of emerging quantum materials.
\end{abstract}

\date{\today}


\maketitle 

\section{Introduction}

FeSe is a paradigmatic model system for unconventional superconductivity that is continuing to attract significant attention due to enhanced superconducting transition temperature $T_c$ in the electron-doped FeSe \cite{Miyata2015high,Wen2016anomalous}, single-layer FeSe/SrTiO$_{3}$ \cite{Wang2012interface,He2013phase}, and the connection of this parent material to possible topological superconductivity in FeSe$_{0.5}$Te$_{0.5}$ \cite{Wang2015topological,Zhang2018observation}. Despite a significant amount of research, the question of fundamental pairing symmetry continues to be debated even in the parent, single crystals of FeSe. In fact, $s$-wave, $d$-wave, as well as mixed symmetries of the order parameter (e.g. $s+i d$) have been proposed for FeSe depending on the specific theoretical or experimental techniques. Unlike most parent compounds of iron-based superconductors that exhibit long-range antiferromagnetic order \cite{de2008magnetic,kamihara2008iron,zhao2008structural,moon2010chalcogen,paglione2010high,qureshi2010crystal}, FeSe does not exhibit long-range magnetic order \cite{mcqueen2009tetragonal,glasbrenner2015effect,wang2015nematicity,wang2016magnetic} but instead is characterized by nematic order due to broken rotational symmetry that coexists with superconductivity \cite{wang2015nematicity,shimojima2014lifting,watson2015emergence,bohmer2017nematicity}.
It has therefore been suggested that superconducting pairing is driven by spin fluctuations that require the sign-changing order parameters such as $s_\pm$ \cite{mazin2008unconventional,ning2010contrasting,hirschfeld2011gap,dai2015antiferromagnetic}. 
There is indeed broad agreement that the superconducting order parameter in FeSe involves a sign change of phase across the Fermi surface, in particular, to minimize the strength of Coulomb repulsion.
However, direct probes of sign-changing order parameter have been challenging. In order to explain the nematic order and strong anisotropic gap in FeSe, the mixed $s + d$ wave scenario has been invoked \cite{livanas2015nematicity,kang2018superconductivity,kushnirenko2018three}. In addition, a transition
to an $s+e^{i\alpha} d$ pairing that breaks the time-reversal symmetry below $T_c$ has been also proposed \cite{kang2018time,islam2021specific}. 

\begin{figure}[t!]
  \centering
  \includegraphics[width=\columnwidth]{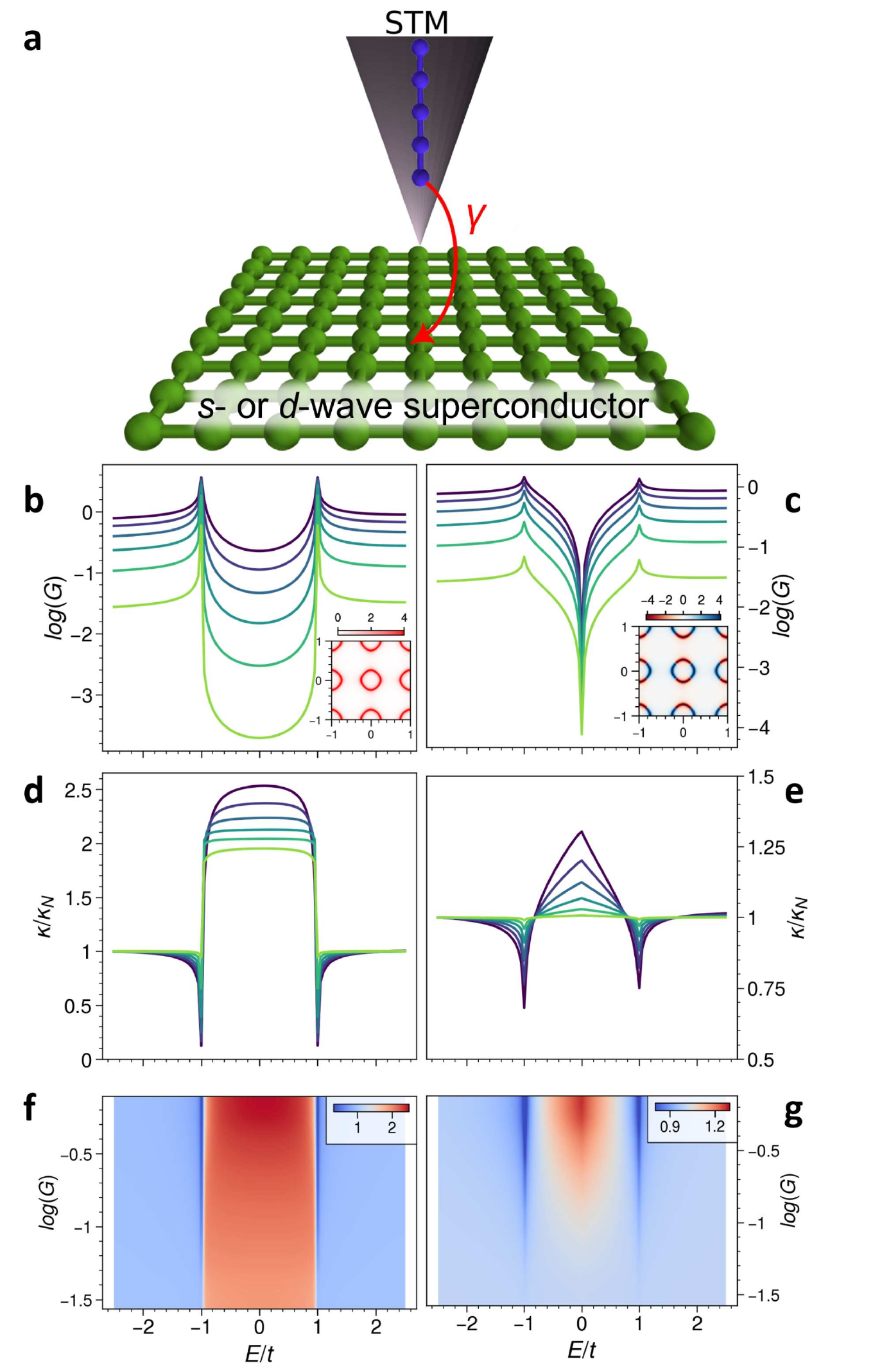}
     \caption{
     \textbf{Illustration of the methodology of Tunneling Andreev Reflection (TAR) and its ability to differentiate superconducting order parameters.} \textbf{a}, Schematic of the transport setup used in the tight-binding modeling of TAR. \textbf{b},\textbf{c}, Conductance spectra for $s$-wave and $d$-wave order parameters as a function of tunneling barrier transparency $\gamma = 0.1$ (light green) $\sim 0.6$ (dark blue). Insets show the corresponding Fermi surfaces. The superconducting gap persists qualitatively unchanged across the whole range of $\gamma$. \textbf{d},\textbf{e}, The corresponding renormalized decay rate ($\kappa/\kappa_{N}$) spectra for the two order parameters. The onset of higher-order Andreev reflection is marked by $\kappa/\kappa_{N} > 2$ for the $s$-wave and $\kappa/\kappa_{N} > 1$ for the $d$-wave, providing key evidence to differentiate between the two symmetries in experiments. \textbf{f},\textbf{g}, Colormaps of the decay rate spectra in \textbf{d} and \textbf{e}, respectively, as a function of conductance log$(G)$ ($y$-axis) and energy $E/t$ ($x$-axis). 
     }
     \label{fig:fig1}
\end{figure}

Notably, angle-resolved photoemission spectroscopy (ARPES) and scanning tunneling microscopy (STM) - as direct probes of electronic structure - continue to provide crucial evidence in support of a specific model of the order parameter. Especially for the STM, quasiparticle interference (QPI) technique has been extensively developed and applied to both cuprates \cite{Hoffman2002,Hanaguri2007,Zou2022} and iron-based superconductors \cite{Chuang2010, Allan2012, Kostin2018} to probe the anisotropic nature of the gap, as well as the symmetry of the order parameter. The phase-corrected quasiparticle interference, in particular, has been used to infer sign-changing order parameter in FeSe and several related materials, relying on the analysis of the scattering of the gap edge states \cite{Chen_2022, Hanaguri_2010, Chen_2019}. However, the connection between QPI signals and symmetry of the order parameter requires sophisticated analysis, that can be further complicated by a narrow energy gap, weak scattering intensity and other factors  \cite{Chen_2022,PhysRevB.92.184513,PhysRevB.97.054519}. Moreover, both QPI and ARPES are quasiparticle probes that are only indirectly sensitive to the superconducting condensate, and neither can effectively tackle inhomogenous and/or spatially localized superconducting states. A notorious challenge, for example, is to distinguish a pseudo-gap V-shaped density of states from a superconducting V-shaped gap (\cite{Fischer_2007} and references therein). The continued uncertainty in the pairing symmetry of even established superconductors strongly motivates the search for techniques that are directly sensitive to the superconducting order in reciprocal space and can provide complementary reinforcing or perhaps contradicting evidence to the established approaches. 

Point-contact Andreev reflection (PCAR) provided crucial evidence in support of the $d$-wave pairing symmetry in cuprate superconductors \cite{PhysRevB.53.2667,PhysRevLett.81.2542,RevModPhys.77.109}, and it can also, at least in principle, differentiate between different symmetries proposed for FeSe \cite{Daghero2011}. In Andreev reflection, the electron is converted into a hole upon reflection from a normal metal-superconductor interface, and a Cooper pair is injected into the superconductor. PCAR measurements on FeSe have revealed the existence of two anisotropic superconducting gaps \cite{Bashlakov2019, PhysRevB.96.094517, PhysRevB.100.104516}, i.e., a picture also consistent with specific heat measurements \cite{lin2011coexistence,chen2017two,jiao2017superconducting}. However, here too numerous experimental conditions need to be satisfied to enable robust identification \cite{Ponomarev2011,PhysRevB.96.094517,Bashlakov2019}, not least of which is the requirement of high-quality directional contacts \cite{Daghero2011} which are difficult to implement particularly in van der Waals solid. The intrinsic challenges of PCAR technique are also exacerbated by the general variability of disorder between different samples, which owing to the small size of the Fermi surface can introduce strong changes in the electronic structure \cite{PhysRevB.100.104516}.  

In this manuscript, we present the first measurement of superconducting FeSe using tunneling Andreev reflection (TAR) - a recently developed methodology that combines the spatial and energy resolution of STM with the direct and quantitative measurement of Andreev reflection \cite{PhysRevResearch.3.033248,Ko2022}. The advantage of the TAR is that it reveals explicit dependency of Andreev current on the tunable tunneling transparency and also uniquely probes higher-order Andreev reflection processes \cite{Ko2022}. Meanwhile, the use of renormalized decay rate as an experimental observable enables robust theory-experiment comparison in TAR. Our measurements unequivocally demonstrate the sign-changing nature of the superconducting order parameter in FeSe, and further ascertain the existence of two superconducting gaps. Given the atomic-scale resolution of STM, we demonstrate that the twin boundaries on FeSe locally suppress superconductivity, explaining their preferential role as vortex pinning centers \cite{PhysRevLett.109.137004,PhysRevX.5.031022}. Overall, our measurements provide new evidence to support a sign-changing pairing symmetry, and also illustrate the new capability of TAR to probe unconventional superconductors via precise control over the tunneling coupling as well as quantitative analysis of lowest- and higher-order Andreev reflection processes.

\begin{figure}[t!]
  \centering
  \includegraphics[width=\columnwidth]{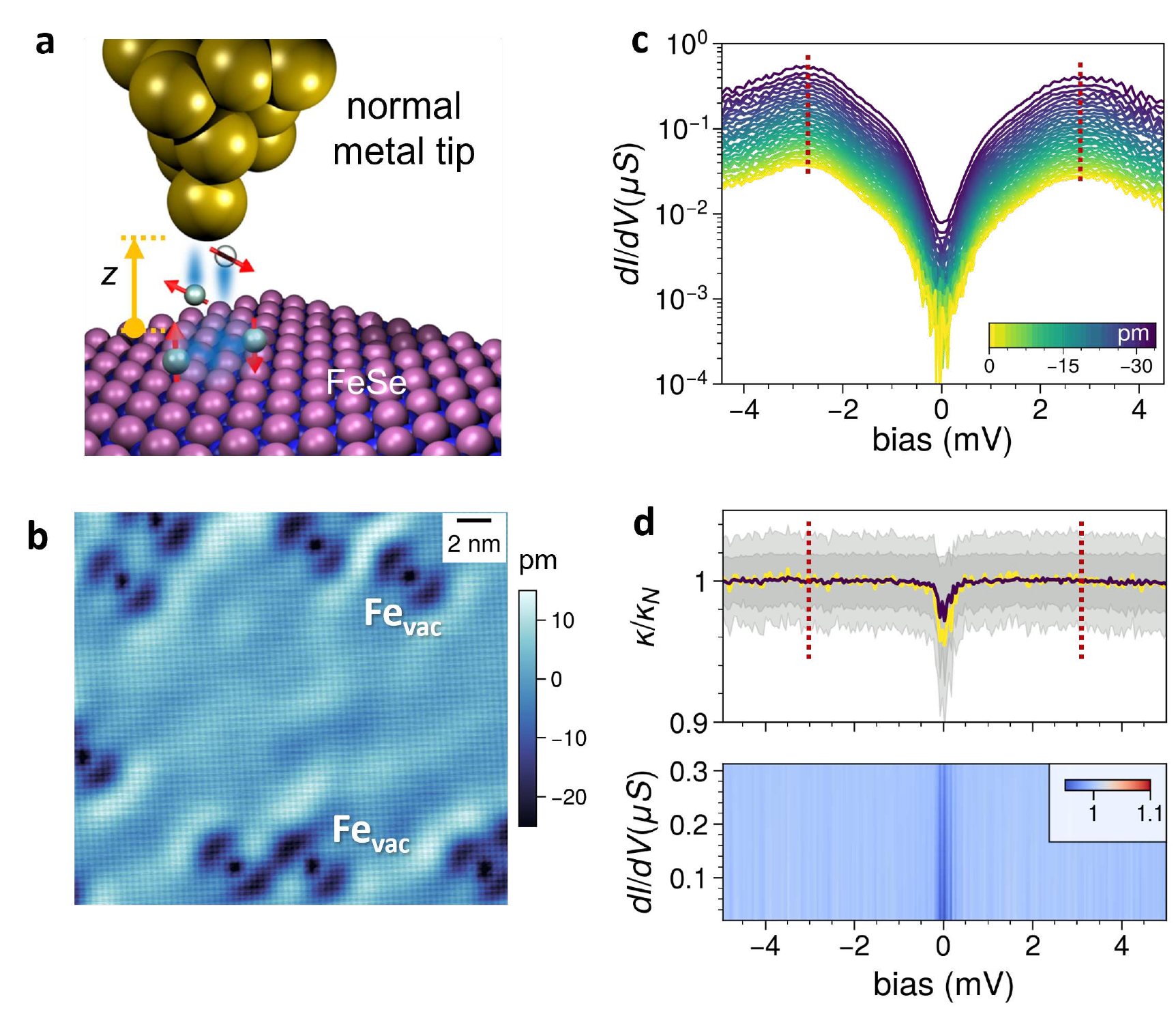}
     \caption
     {
     \textbf{Schematic of experimental measurement of tunneling Andreev reflection.} \textbf{a}, Experimental setup for conductance and decay rate spectra. The spectra are measured between a metallic tip and superconducting FeSe as a function of varying tunneling conductance controlled by tip-surface separation $z$. \textbf{b}, STM image of atomically-flat FeSe surface with finite density of naturally occurring Fe-vacancies (setpoint bias = -10 mV, $I = 50$ pA). \textbf{c}, Conductance spectroscopy displaying superconducting gap as a function of varying $z$ in the range from 0 to -34 pm (negative values correspond to closer proximity to the surface). \textbf{d}, Decay rate spectra obtained from \textbf{c} and its corresponding colormap view (bottom half). Within the errorbars (grey area), $\kappa/\kappa_{N}$ equals unity across the whole superconducting gap, directly witnessing the presence of sign-changing order parameter. The errorbars were estimated from Bayesian fitting of the 2nd-order polynomial function to the $z$-dependency of $dI/dV$ at each tunneling energy in \textbf{c}. Red dashed lines mark approximate edges of the superconducting gap. }
     \label{fig:fig2}
\end{figure}

\begin{figure}[t!]
  \centering
  \includegraphics[width=\columnwidth]{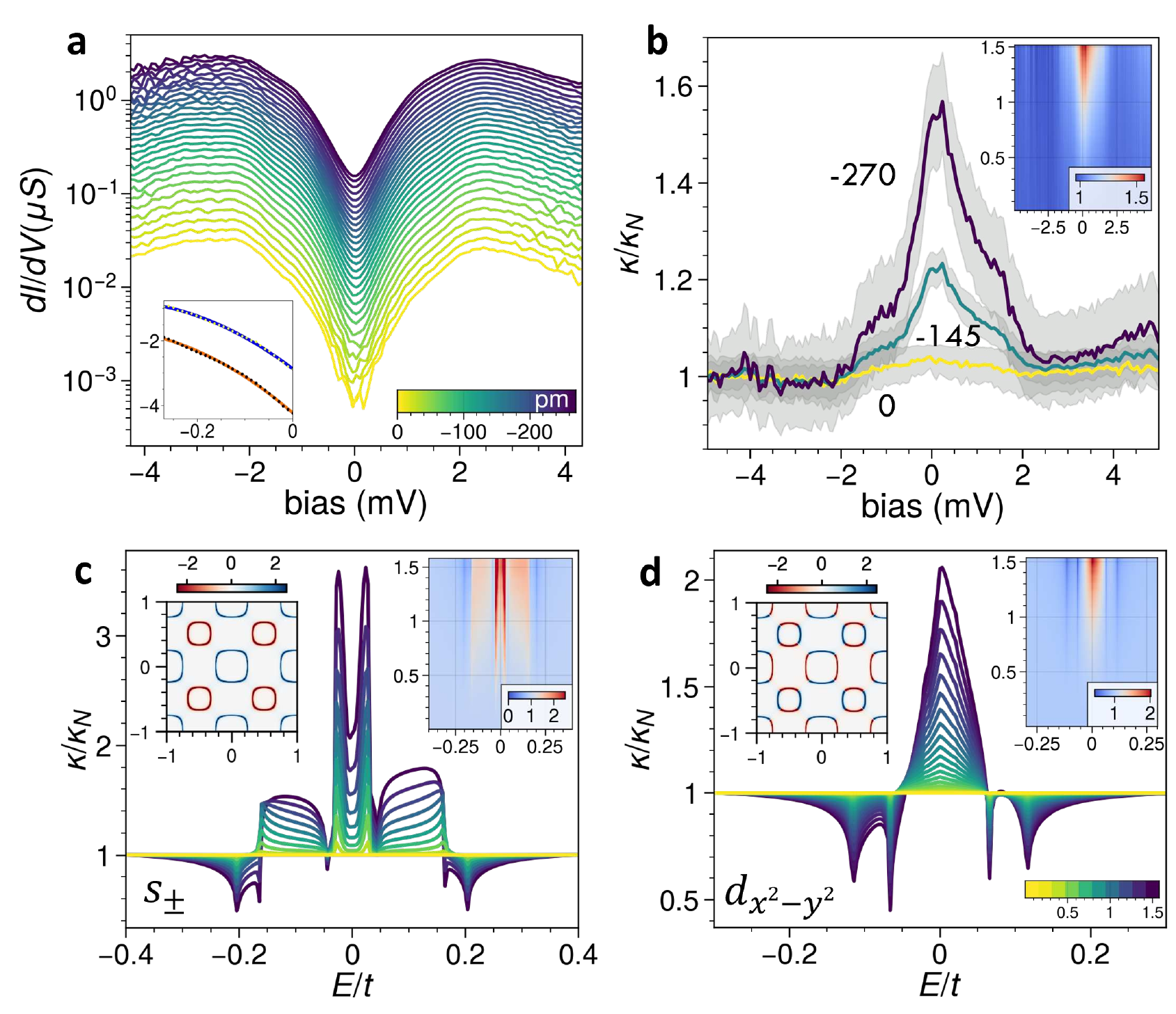}
     \caption{
     \textbf{Emergence of higher-order Andreev reflection at increased tunneling conductance.} \textbf{a}, Measured $dI/dV$ spectra as a function of decreasing $z$ by up to -270 pm. The inset shows individual $z$-dependency ($x$-axis) of $dI/dV$ ($y$-axis) measured at energies of 0 mV (red) and -2 mV (blue) \textbf{b}, Decay rate spectra extracted from \textbf{a} at 0, -140 and -270 pm. Higher-order Andreev reflection is emerging as a peak centered at zero bias with a maximum $\kappa/\kappa_{N}$ value of 1.6. The colormap inset clearly shows transition from suppressed to finite Andreev reflection as the tunneling conductance increases. \textbf{c},\textbf{d}, Tight-binding modeling of higher-order Andreev reflection for $s_\pm$ and $d_{x^2-y^2}$ order parameters with a two-gap structure. The colors of individual spectra correspond to tunneling transparency $\gamma$ changing from 0.01 to 1.5. The insets show the Fermi surface (left) and colormap (right) of decay-rate spectra, with regions of higher-order reflection marked by $\kappa/\kappa_{N} > 1$ (pink and red).}
     \label{fig:fig3}
\end{figure}

\section{Experiment \& Results}

The key distinction of TAR is the choice of observable to be the decay rate of tunneling conductance - a distinct departure from conventional Andreev spectroscopy including its proposed extension toward STM \cite{Sukhachov2022}, all of which rely on the conductance itself. The decay rate fundamentally probes the order of he order of the tunneling process responsible for tunneling, enabling detection of lowest and higher order Andreev reflections \cite{Ko2022}.  Subsequently, as shown in the following, the dependency of Andreev reflection on the transparency of the tunneling junction can be measured by TAR and compared to a variety of theoretical models directly. By contrast, in point-contact measurements, the transparency is generally a specific and unknown experimental value, which dramatically complicates the degree by which pairing symmetry can be inferred \cite{Lee2015proceedings, Daghero2011}. 

To illustrate the fundamental aspects of TAR, we first present tight-binding simulations of the Andreev reflection in a tunneling contact contrasting sign-changing
and non-sign changing order parameters (Fig.~\ref{fig:fig1}a). Here, the tunneling conductance of a metal-superconductor contact is calculated as a function of transmission coefficient $\gamma$ of the tunneling contact for an $s$-wave (Fig.~\ref{fig:fig1}b) and $d$-wave (Fig.~\ref{fig:fig1}c) Bogoliubov-de-Gennes Hamiltonian. The detection of Andreev current can be extracted via a renormalized decay rate of tunneling conductance  $\kappa/\kappa_{N}$ (Fig.~\ref{fig:fig1}d-g) \cite{Ko2022}. By virtue of internal normalization, $\kappa/\kappa_{N}$ is particularly well-suited for theory-experiment matching, and it reduces the effect of numerous systematic and random unknowns in both calculation and, most importantly, experimental measurement of the tunneling conductance. 

For a conventional $s$-wave superconductor, the tunneling current within the superconducting gap is fully suppressed at $T = 0$ K, except for Andreev reflection, which creates finite conductance even in the middle of the gap as seen in Fig.~\ref{fig:fig1}b. At small tunneling conductance, the decay rate due to the Andreev current is twice that of the normal current at the Fermi level. The superconducting gap is then registered by a characteristic 1 to 2 transition of $\kappa/\kappa_{N}$ across the gap as shown in Figure \ref{fig:fig1}d. However, the maximum value of $\kappa/\kappa_{N}$ itself depends on the tunneling conductance, increasing beyond 2 when $\gamma > 0.3$  (Fig.~\ref{fig:fig1}d,f). This increase of $\kappa/\kappa_{N}$ signals the regime of higher-order Andreev reflection. Its additional signature is that $\kappa/\kappa_{N} < 1$ (resonant enhancement) at the edges of the gap as seen in Fig.~\ref{fig:fig1}d. The conceptual origin of the higher-order processes can be rationalized by considering tunneling contact perturbatively, i.e., the dependency of tunneling conductance on increasing powers of the transmission coefficient $\gamma$. Under the conditions of weak coupling (small $\gamma$), only lowest order terms will contribute to the tunneling conductance, which corresponds to single electron tunneling ($\propto \gamma$) and two-particle Andreev reflection ($\propto \gamma^{2}$) in normal and superconducting states, respectively. However, when the tunneling coupling increases, the value of higher-order Andreev reflection terms (e.g. $\propto \gamma^{4}$) become non-negligible, and this is directly reflected in the value of $\kappa/\kappa_{N}$ increasing beyond $2$. 

\begin{figure}[t!]
  \centering
  \includegraphics[width=\columnwidth]{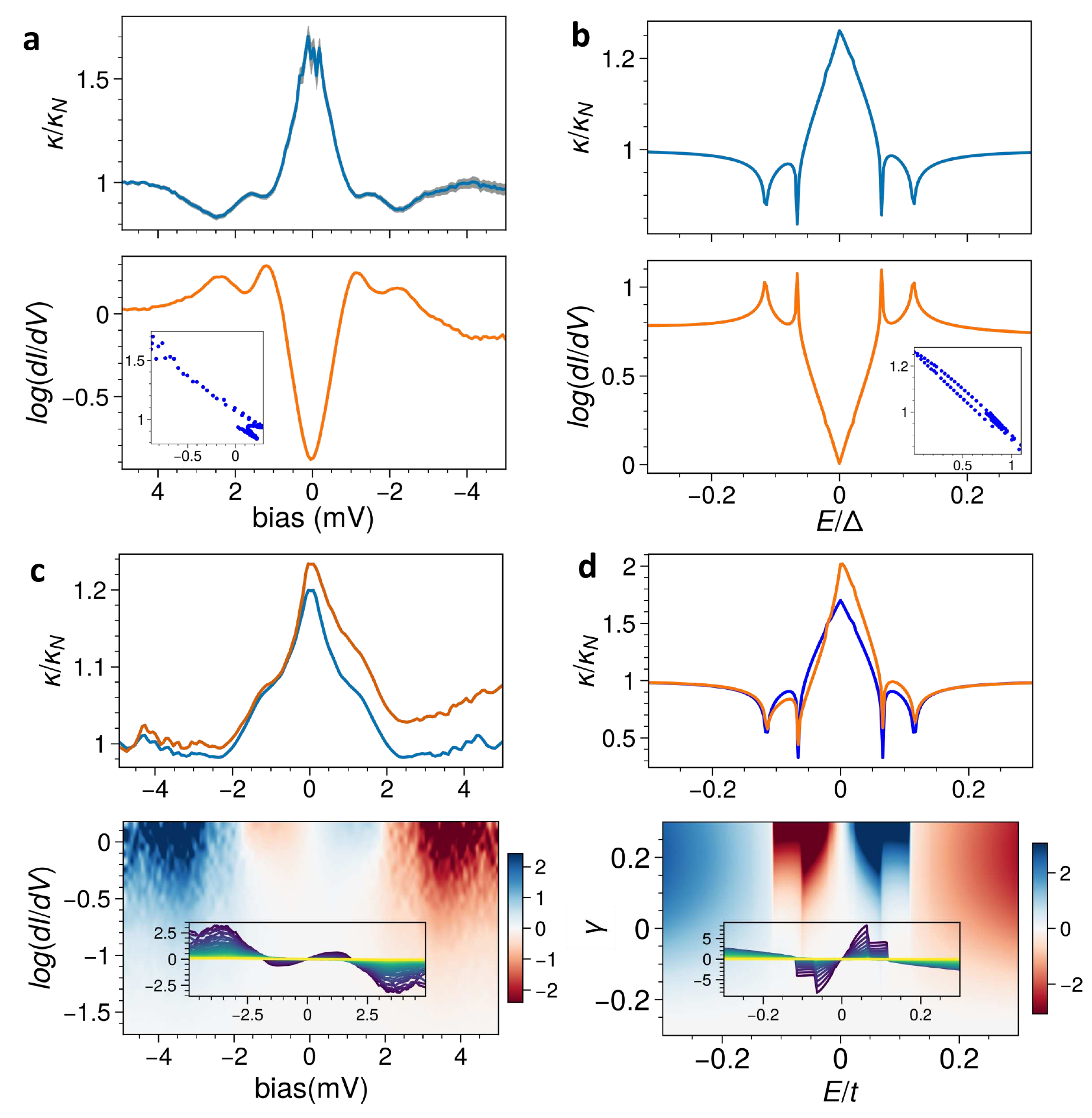}
     \caption{
     \textbf{Comparison of measured and calculated decay rate spectra for the $d_{x^2-y^2}$ order parameter.} \textbf{a}, The correspondence and anticorrelation between the measured decay rate spectra (top) and the conductance spectra (bottom). Inset shows the nearly linear anticorrelation between decay rate and conductance measured on FeSe surface. \textbf{b}, The correspondence and anticorrelation between decay rate and conductance spectra from tight-binding modeling with a sign-changing order parameter, closely reminiscent to the measurements in \textbf{a}. \textbf{c}, Asymmetry of decay rate spectra manifested by comparison between the decay rate spectrum measured at 42 $\mu$S (orange) and its symmetrized form (blue) (top panel). Bottom panel shows the asymmetry of the conductance spectra as a function of the tunneling conductance. The asymmetry is registered as relative increase of the density of states below Fermi level at energies outside the superconducting gap, and oppositely, a relative decrease of the density of states within the superconducting gap region. \textbf{d}, Tight-binding modeling that qualitatively reproduces all the asymmetries observed experimentally, both in the decay rate and conductance spectra.
     }
     \label{fig:fig4}
\end{figure}

For a sign-changing $d$-wave order parameter with a nodal gap, TAR spectroscopy is qualitatively different from that of the isotropic $s$-wave (Fig.~\ref{fig:fig1}c,e). The nodal quasiparticle tunneling completely suppresses lowest-order Andreev reflection, which is seen as $\kappa/\kappa_{N}$ = 1 across the gap at $\gamma$ = 0.1 (Fig.~\ref{fig:fig1}e). This process occurs because Andreev reflection decays twice as fast as single-particle tunneling, leaving the latter channel dominant at small tunneling conductance. However, the contribution of higher-order Andreev reflection still emerges with increased tunneling conductance. Indeed, $\kappa/\kappa_{N}$ becomes enhanced in the middle of the gap and again suppressed at the edges (Fig.~\ref{fig:fig1}e,g). The net result is a central peak structure with height rapidly increasing with increased tunneling conductance. Therefore the combination of lowest- and higher-order Andreev reflection phenomena accessible in tunneling junctions, complementary to the traditional conductance spectroscopy, enables a considerably robust test for both the superconductivity itself and the existence of a nodal structure in the superconducting gap. 

Figure \ref{fig:fig2}a shows the schematic of the TAR experiment applied to FeSe. The transparency of the contact is controlled by varying the separation between the metal tip and atomically flat FeSe surface (Fig.~\ref{fig:fig2}b). The $\kappa/\kappa_{N}$ spectra are obtained from $z$-dependent tunneling spectra of the superconducting gap (Fig.~\ref{fig:fig1}c) as described in \cite{Ko2022}. There are two questions posed for TAR: (1) does $\kappa/\kappa_{N}$ deviate from a value of 1 across the gap; and (2) are there signatures of the higher-order Andreev reflection beyond some value of tunneling conductance? Note that the crossover value of tunneling conductance that separates lowest- and higher-order regimes of Andreev reflection is not known
a priori. The crossover value shall depend on the exact tip configuration that has substantial variability across experiments, not in small part due to largely unavoidable interactions with the chalcogenide layer at close proximity. 

Figure \ref{fig:fig2}d shows the decay rate spectroscopy for the range of tunneling conductance up to 0.5 $\mu$S (measured at 4.5 mV bias). As seen in Fig.~\ref{fig:fig2}d, $\kappa/\kappa_{N}$ maintains a value of 1 across the whole gap within the statistical uncertainty (The error bars were obtained from Bayesian fitting of the 2nd-order polynomial to the $z$-dependence of $dI/dV$). This result is qualitatively different from plateau of $\sim$2 observed in a conventional superconductor (Fig.~\ref{fig:fig1}d) \cite{Ko2022}. Suppression of Andreev reflection in the tunneling regime already indicates the high likelihood of a sign-changing gap structure, much more directly than the shape of the tunneling spectroscopy itself. 

Meanwhile, extending the measurements to 2-3 $\mu$S begins to notably grow $\kappa/\kappa_{N}$ in the middle of the gap, developing a peak-like structure with pronounced shoulders and a maximum observed value of about 1.6 in the middle of the gap (Fig.~\ref{fig:fig3}a,b). The experimental spectra of higher-order Andreev reflection exhibit some expected amount of experimental variability but maintain their overall shape, as shown in Fig.~S1. Already at this stage, we can make a direct comparison to the results of tight-binding modeling with the sign-changing order parameter symmetries for FeSe (Fig.~S3), specifically the extended $s$-wave shown in Fig.~\ref{fig:fig3}c and nodal $d$-wave in Fig.~\ref{fig:fig3}d, calculated for two-pocket Fermi surface consistent with FeSe. 

The Fermi surface of FeSe constitutes hole Fermi pockets at $\Gamma$ point, and electron Fermi pockets at M point \cite{qureshi2010crystal,lebegue2007electronic,fanfarillo2016orbital} (Fig.~S2). For both order parameters, Andreev reflection is suppressed at small tunneling conductance (yellow lines in Fig.~\ref{fig:fig3}c,d), in close agreement with Fig.~\ref{fig:fig2}d. Notably, this is also true for the fully gapped $s_\pm$ order parameter, where the nodal lines do not intersect Fermi surface pockets. The Andreev
signal, in this case, is suppressed because of destructive interference between opposite-signed pockets of the Fermi surface - which is uniquely probed by TAR. Meanwhile, calculations for both $s_\pm$ and $d$-wave exhibit higher-order Andreev reflection, albeit with a different spectral signature. The experimental signal in Fig.~\ref{fig:fig3}b exhibits a peak-and-shoulder structure concentrated around the middle of the gap, similar to the sign-changing $d$-wave order parameter in \ref{fig:fig3}d. The extended $s$-wave in Fig.~\ref{fig:fig3}c exhibits larger $\kappa/\kappa_{N}$ across the full width of the superconducting gap and even slight enhancement at the gap edges. It is worth noting that the average gap is zero when integrated in the Fermi surface for $d_{x^2-y^2}$ order, i.e, $\langle \Delta_{d_{x^2-y^2}} (\mathbf k) \rangle_{FS} = 0$, whereas $\langle \Delta_{s_{\pm}} (\mathbf k) \rangle_{FS} \ne 0$ for $s_{\pm}$ order. For the transport setup of our tight-binding modeling, this gap average determines the behavior of $\kappa/\kappa_{N}$, accounting for the differences between the two orders. Comparison with other superconducting orders is included in the supplemental information (Fig.~S4).

Two additional characteristics of the higher-order Andreev reflection allow us to establish further agreement between modeling and experiment, and also to directly infer the existence of a two-gap superconducting structure. It is clear from the modeling results in Figs. \ref{fig:fig1}, \ref{fig:fig3} and Fig.~S4 that $\kappa/\kappa_{N}$ is reduced below 1, i.e. tunneling conductance is resonantly enhanced, at the edges of the superconducting gap, including each gap of the multigap structure. Although this effect is necessarily broadened by lifetime and thermal effects, we directly observed it in decay rate spectroscopy on Pb(110) surface \cite{Ko2022}. In the case of FeSe, depending on specific imaging conditions and particularly at large tunneling conductance, the conductance spectra clearly manifest the inner-gap structure, with gap positions at $\pm 1.1$ meV (Fig.~\ref{fig:fig4}a, orange curve). The energy of the inner gap is consistent with previously published results \cite{PhysRevB.96.094517}. For the tunneling conditions corresponding to these spectra, the shoulder structure of higher-order Andreev reflection from Fig.~\ref{fig:fig3}b becomes more pronounced, with now three clear peaks in $\kappa/\kappa_{N}$ within $\pm 2.5$ meV range. Moreover, the dips separating these peaks coincide within measurement accuracy with the peaks in the $dI/dV$ spectrum in near exact similarity with the simulated results shown in Fig.~\ref{fig:fig4}b. The experimentally observed resonant enhancement at the gap edges is less pronounced than in the calculations, which is natural considering the broadening effects. However, the observation directly points toward a two-gap structure of the superconducting gap, confirming both earlier results and providing another point of agreement to ascertain the origins of experimentally measured $\kappa/\kappa_{N}$ to be higher-order Andreev reflection.

\begin{figure}[t!]
  \centering
  \includegraphics[width=\columnwidth]{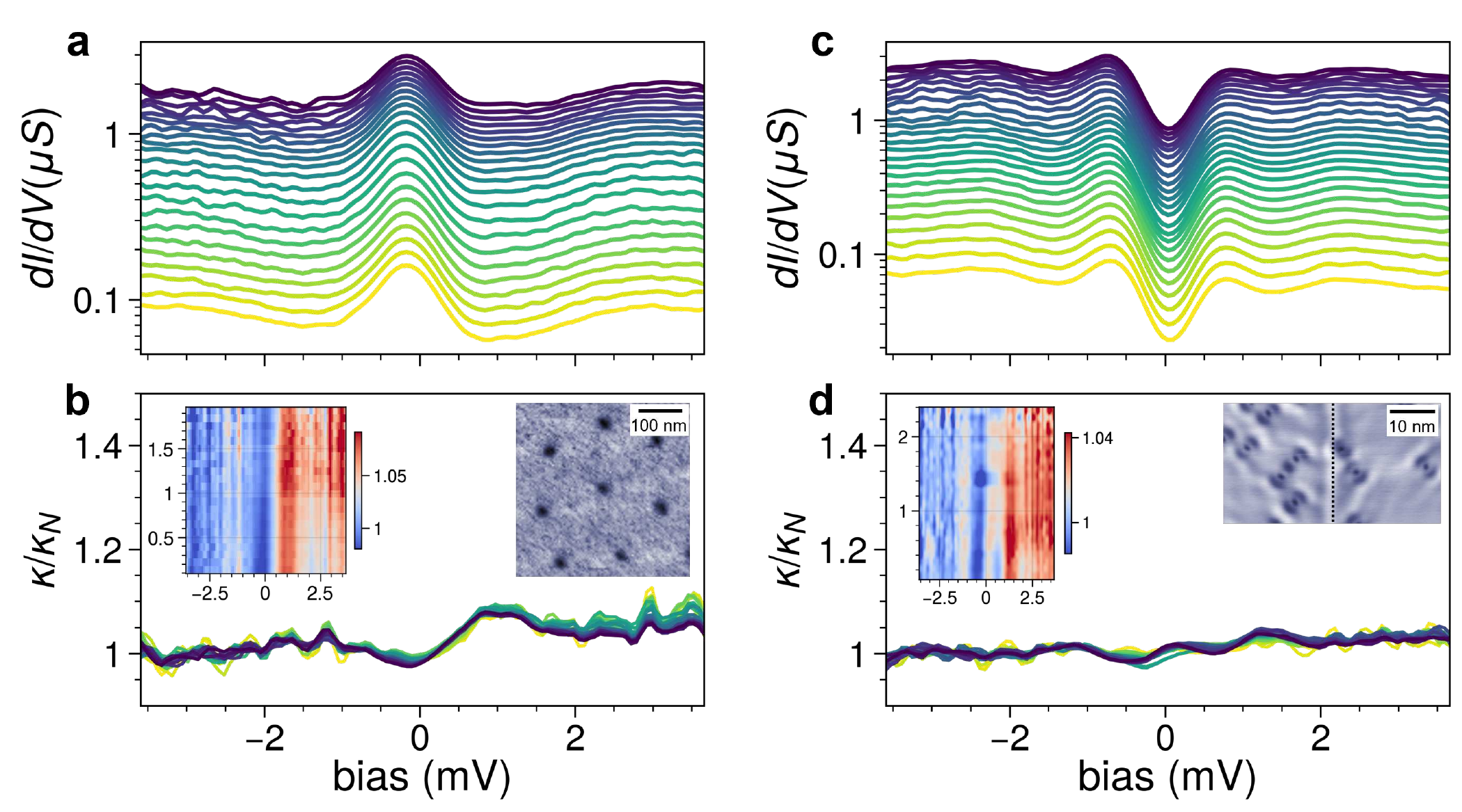}
     \caption{
     \textbf{Tunneling Andreev reflection of topological defects on FeSe.} Conductance spectra (\textbf{a},\textbf{c}) and their corresponding decay rate spectra (\textbf{b},\textbf{d}) acquired at the center of the superconducting vortex (\textbf{a},\textbf{b}) and the nematic twin boundary (\textbf{c},\textbf{d}). $\kappa/\kappa_{N}$ spectra show no significant deviation from unity for both vortex and twin boundary, with a possible expectation of slightly asymmetric feature at the vortex. Insets of \textbf{b},\textbf{d} show correspondingly the colormaps of the decay-rate spectra (left) and the representative STM images of the surface areas of FeSe containing the topological defects (right)(imaging setpoint bias = -2.5 mV, $I = 100$ pA for \textbf{b}; bias = -10 mV, $I = 50$ pA for \textbf{d}). The range of tunneling conductance probed for both entities is comparable to or even exceeding that for bare surface in Fig.~\ref{fig:fig3}b.
     }
     \label{fig:fig5}
\end{figure}

Another property of the decay rate spectra is the asymmetry around Fermi level, frequently observed in the experiment as shown in Fig.~\ref{fig:fig4}c (orange line). In particular, $\kappa/\kappa_{N}$ outside of the superconducting gap (larger than 2.5 meV energy) is higher at energies above the Fermi level than below. This effect can be numerically confirmed by symmetrizing the measured $dI/dV$ around Fermi level prior to calculating $\kappa/\kappa_{N}$, resulting in the blue line in Fig.~\ref{fig:fig4}c. Note that the overall peak structure is not strongly affected. Fig.~\ref{fig:fig4}c bottom panel also shows the asymmetric part of the $dI/dV$ spectrum around Fermi level, which gives rise to the asymmetry in $\kappa/\kappa_{N}$. The asymmetry is also observed in tight-binding simulations (Fig.~\ref{fig:fig4}d). Here, the asymmetry of $\kappa/\kappa_{N}$ directly stems from the asymmetry of the electronic structure in the normal state. Interestingly, this asymmetry in the spectra is manifested more strongly closer to the contact. This behavior can be readily rationalized from the perturbative point of view, where a small asymmetry in a low-order process $\alpha(\omega)/\alpha(-\omega) \approx 1 - \epsilon$ becomes stronger for higher order processes $\alpha(\omega)^N/\alpha(-\omega)^N \approx 1 - N\epsilon$.

The above measurements all provide strong new evidence for the unconventional sign-changing superconducting order parameter in FeSe with two gaps around 1.1 and 2.5 meV energies, one in the $\Gamma$ pocket and one in the M pocket. Moreover, TAR spectroscopy is more consistent with $d_{x^2-y^2}$ order parameter symmetry, rather than $s_{\pm}$ within the accuracy of present measurements and transport modeling. Reducing the broadening effects, for example, by measurements at progressively lower temperatures will help make this distinction even more clear in the future (Fig.~S5).

We now apply the above methodology to two kinds of topological defects that are characteristic of superconducting FeSe - an Abrikosov vortex and a nematic twin boundary (Fig.~\ref{fig:fig5}). Because TAR relies on properties of the atomic-scale junction, it is uniquely poised to explore spatially inhomogenous and/or localized superconductivity. Since the superconducting order parameter is suppressed in the vortex center, the ansatz for $\kappa/\kappa_{N}$ is that its value stays close to 1, indicating weak energy dependence of normal tunneling in the $\pm 2$ meV window around the Fermi level. The experimental decay rate spectra across the range of tunneling conductance comparable to Fig.~\ref{fig:fig2} and \ref{fig:fig3} indeed reveal a value that is very close to unity (Fig.~\ref{fig:fig5}b). This observation is even more reassuring since the vortex-bound state is clearly observed in the $dI/dV$ spectra (Fig.~\ref{fig:fig5}a), which is perhaps responsible for the slight modulation of $\kappa/\kappa_{N}$ around Fermi level. For reference, the spectra of the normal state FeSe above the superconducting gap reveal hardly any features (Fig.~S1d), further ascertaining that the tunneling regime itself is robust in the probed range and the analysis is largely free from numerical artifacts.

Interestingly, the spectra taken above the nematic twin boundaries likewise reveal no significant deviation of $\kappa/\kappa_{N}$ from unity across the whole energy range of the superconducting gap (Fig.~\ref{fig:fig5}d). The role of twin boundaries in modulating the superconductivity of FeSe is another somewhat controversial point. From prior STM measurements, it is known that the twin boundaries exhibit a gap structure that is narrower than the FeSe gap value of $\pm 2.5$ meV, and this is also clearly seen in our $dI/dV$ spectra in Fig.~\ref{fig:fig5}c. It was suggested that superconductivity is therefore reduced, but still present along the boundaries \cite{PhysRevLett.109.137004} and may even be used as evidence of time-reversal symmetry breaking order parameter \cite{PhysRevX.5.031022}. However, given our results from TAR, we propose that superconductivity is completely suppressed at the twin boundary, at least above our measurement temperature of 1.2 K. Indeed, there is no signature of lowest- or higher-order Andreev reflection in the spectra in Fig.~\ref{fig:fig5}d, which is dramatically different from the properties of the adjacent FeSe regions. The twin boundary, therefore, acts as an extended defect, suppressing the unconventional superconducting order in FeSe due to the pair breaking effect. Any type of local defect that creates momentum scattering will average out the sign-changing gap in reciprocal space, giving rise to in-gap modes locally quenching the superconducting order. This mechanism is analogous to the quenching of superconductivity in conventional superconductors by magnetic impurities \cite{RevModPhys.78.373}. The question of the origin of the observed gap in $dI/dV$ spectra on twin boundaries (Fig.~\ref{fig:fig5}c) remains open for now. One possibility is that this is the pseudogap in the density of states that stems from the normal state and is not directly related to superconductivity \cite{Fischer_2007}. The non-superconducting twin boundary is also very consistent with the behavior of superconducting vortices in FeSe, which are preferentially attracted to the boundary \cite{PhysRevLett.109.137004,PhysRevX.5.031022}.

\section{Conclusions}

Here we have presented the first measurements of tunneling Andreev reflection on FeSe superconductor. The unique ability to probe both lowest- and higher-order Andreev reflection processes in tunneling measurements enables probing the sign-changing multigap structure of the superconducting order parameter, while simultaneously having access to the local density of states from conductance spectroscopy and atomic-scale resolution. The lowest-order Andreev reflection is suppressed on FeSe within the measurement accuracy, while higher-order processes reveal a characteristic peak-like shape with pronounced shoulders. Through detailed comparison to tight-binding modeling of the many candidate order parameters, our results provide the first direct measurement of the sign-changing nature of the superconducting order parameter with atomic spatial resolution. Meanwhile, the twin boundary exhibits no detectable Andreev reflection, demonstrating the completely suppressed superconductivity along the boundary above the lowest measurement temperature of 1.2 K. The observation implies that the twin boundary can be considered as a narrow geometric region of metallic conductance, acting as an intrinsic Josephson junction coupling adjacent regions of superconducting FeSe. 

The measurements provide additional evidence to probe the fundamental origins of superconducting pairing in FeSe, and set the stage for quantitative microscopic characterization of the phase diagram of FeSe as a function of doping and Te substitution. Furthermore, by taking advantage of the tunability of tunneling coupling, the methodology of TAR enables new insight into pairing mechanisms of unconventional superconductors and is likely to provide valuable insight into the properties of emerging quantum materials, including exotic superconductivity for quantum information science.

\section{Methods}

The experiments were performed in a SPECS JT-STM operated under ultrahigh vacuum condition ($< 10^{-10}$ mbar) and base temperature of 1.2 K. The STM tips were confirmed to be metallic on the Cu(111) single crystal, and further conditioned by pulsing and soft crashes if necessary. FeSe surfaces were prepared by cleaving single crystals in high vacuum ($\sim$$10^{-8}$ mbar) and immediately transferring them to ultrahigh vacuum conditions. The spectra of differential tunneling conductance, $dI/dV$, were obtained by conventional lock-in amplifier method with modulation voltage 50 $\sim$ 80 $\mu$V and modulation frequency 735 Hz. The decay constant $\kappa$ was measured by taking $dI/dV$ spectra for varying tip height $z$ and then numerically differentiate to attain $\kappa = \frac{1}{dI/dV} \cdot \frac{d(dI/dV)}{dz}$, following procedures described prior \cite{Ko2022}.

\textbf{Acknowledgements}:
Experimental measurements were supported by the U.S. Department of Energy, Office of Science, Materials Sciences and Engineering Division (W.K., P.M.). Experiments were carried out as part of the user project at the Center for Nanophase Materials Sciences, Oak Ridge National Laboratory, which is a US Department of Energy Office of Science User Facility. J.L.L. acknowledges the computational resources provided by the Aalto Science-IT project, and the financial support from the Academy of Finland Projects No. 331342 and No. 336243, and the Jane and Aatos Erkko Foundation.

\bibliography{biblio}

\end{document}